\documentstyle[preprint,prl,aps,amsfonts,epsfig,latexsym]{revtex}

\begin{document}

\newcommand{\1}{{\bf \scriptstyle 1}\!\!{1}}
\newcommand{\I}{{\rm i}}
\newcommand{\p}{\partial}
\newcommand{\D}{^{\dagger}}
\newcommand{\bl}{{\bf l}}
\newcommand{\bx}{{\bf x}}
\newcommand{\bq}{{\bf q}}
\newcommand{\bk}{{\bf k}}
\newcommand{\bv}{{\bf v}}
\newcommand{\bp}{{\bf p}}
\newcommand{\bu}{{\bf u}}
\newcommand{\bA}{{\bf A}}
\newcommand{\bB}{{\bf B}}
\newcommand{\bK}{{\bf K}}
\newcommand{\bL}{{\bf L}}
\newcommand{\bP}{{\bf P}}
\newcommand{\bQ}{{\bf Q}}
\newcommand{\bS}{{\bf S}}
\newcommand{\bH}{{\bf H}}
\newcommand{\bI}{{\bf I}}
\newcommand{\balpha}{\mbox{\boldmath $\alpha$}}
\newcommand{\bsigma}{\mbox{\boldmath $\sigma$}}
\newcommand{\bSigma}{\mbox{\boldmath $\Sigma$}}
\newcommand{\bOmega}{\mbox{\boldmath $\Omega$}}
\newcommand{\bomega}{\mbox{\boldmath $\omega$}}
\newcommand{\bpi}{\mbox{\boldmath $\pi$}}
\newcommand{\bphi}{\mbox{\boldmath $\phi$}}
\newcommand{\bnabla}{\mbox{\boldmath $\nabla$}}
\newcommand{\bmu}{\mbox{\boldmath $\mu$}}
\newcommand{\bepsilon}{\mbox{\boldmath $\epsilon$}}

\newcommand{\iLambda}{{\it \Lambda}}
\newcommand{\cA}{{\cal A}}
\newcommand{\cD}{{\cal D}}
\newcommand{\cF}{{\cal F}}
\newcommand{\cL}{{\cal L}}
\newcommand{\cH}{{\cal H}}
\newcommand{\cI}{{\cal I}}
\newcommand{\cO}{{\cal O}}
\newcommand{\cR}{{\cal R}}
\newcommand{\cU}{{\cal U}}
\newcommand{\cT}{{\cal T}}

\newcommand{\be}{\begin{equation}}
\newcommand{\ee}{\end{equation}}
\newcommand{\bea}{\begin{eqnarray}}
\newcommand{\eea}{\end{eqnarray}}
\newcommand{\beqa}{\begin{eqnarray*}}
\newcommand{\eeqa}{\end{eqnarray*}}
\newcommand{\nn}{\nonumber}
\newcommand{\DD}{\displaystyle}

\newcommand{\ba}{\left[\begin{array}{c}}
\newcommand{\baa}{\left[\begin{array}{cc}}
\newcommand{\baaa}{\left[\begin{array}{ccc}}
\newcommand{\baaaa}{\left[\begin{array}{cccc}}
\newcommand{\ea}{\end{array}\right]}

\draft

\title{Quantum Spin Dynamics in Molecular Magnets}
\author{Michael N. Leuenberger, Florian Meier, and Daniel Loss}
\address{Department of Physics and Astronomy, University of Basel, 
Klingelbergstrasse 82, 4056 Basel, Switzerland}
\date{\today}

\maketitle

\hspace*{-\parindent}{\bf Summary.}
The detailed theoretical understanding of quantum spin dynamics in 
various molecular magnets is an important step on the roadway to 
technological applications of these systems. Quantum effects in both 
ferromagnetic and antiferromagnetic molecular clusters are, by now, 
theoretically well understood. Ferromagnetic molecular
clusters allow one to study the interplay of incoherent quantum tunneling
and thermally activated transitions between states with different spin
orientation. The Berry phase oscillations found in Fe$_8$ are signatures
of the quantum mechanical interference of different tunneling paths. 
Antiferromagnetic molecular clusters are promising candidates
for the observation of coherent quantum tunneling on the mesoscopic scale.
Although challenging, applications of molecular magnetic clusters for 
data storage and quantum data processing are within experimental reach already
with present day technology.\\

\hspace*{-\parindent}{\bf Introduction}\\
Molecular magnets have attracted considerable interest recently because
of their potential for data storage and data 
processing~\cite{leuenberger:01}. In addition to possible
future technological applications, molecular magnets are also interesting from
an academic point of view because they show quantum effects
on the mesoscopic scale~\cite{leggett:95} in the form of tunneling of
magnetization. In the following, we review some of our theoretical work
on quantum spin dynamics in molecular magnets. 

Ferromagnetic molecular 
magnets such as Mn$_{12}$ and Fe$_8$ show incoherent tunneling of the
magnetization~\cite{korenblit:78,hemmen:86,enz:86,chudnovsky:88} 
and allow one to study the interplay of thermally
activated processes and quantum tunneling. 
The spin tunneling leads to two effects. Firstly, the magnetization relaxation 
is accelerated whenever spin states of opposite direction
become degenerate due to the variation
of the external longitudinal magnetic field~\cite{Friedman,Thomas,LL1Mn12,LL2Mn12,LL3Mn12}. 
Secondly, the spin acquires a Berry phase during the tunneling process, which leads
to oscillations of the tunnel splitting
as a function of the external transverse magnetic 
field~\cite{Wernsdorfer:Science,Wernsdorfer:EPL,LLFe8,LLBerry}.

Due to the strong quantum spin dynamics induced by antiferromagnetic 
exchange interaction~\cite{barbara:90,krive:90,awschalom:92,gider:95}, 
antiferromagnetic molecular magnets such as ferric 
wheels belong to the
most promising candidates for the observation of coherent quantum
tunneling on the mesoscopic 
scale~\cite{chiolero:98,meier:01,meier:01b,honecker:01}. 
In contrast to incoherent tunneling, in quantum coherent tunneling 
spins tunnel back and forth between energetically degenerate
configurations at a tunneling rate which is 
{\it large} compared to the decoherence rate. The detection of coherent 
quantum tunneling is more challenging in antiferromagnetic molecular magnets
than in ferromagnetic systems, but is feasible with present day experimental 
techniques. 

Understanding the properties of molecular magnets is only a first step on the
roadway to technological applications. A possible next step will be the
preparation and control of a well defined single-spin quantum state of a molecular cluster.
Although challenging, this task appears feasible with present day experiments
and would allow one to carry out quantum computing with molecular 
magnets~\cite{leuenberger:01}. 
The idea is to use the Grover quantum search algorithm~\cite{Grover} to read-in and decode
information stored in the phases of a single-spin state. \\


\hspace*{-\parindent}{\bf Spin tunneling in Mn$_{12}$-acetate}\\
The magnetization relaxation of crystals and powders made of molecular magnets Mn$_{12}$ has attracted much recent interest since several experiments \cite{Paulsen,Paulsen2,Novak,Sessoli,Novak2} have indicated unusually long relaxation times as well as increased relaxation rates\cite{Friedman,Thomas,Hernandez} whenever two spin states become degenerate in response to a varying longitudinal magnetic field $H_z$. According to earlier suggestions\cite{Barbara,Novak} this phenomenon has been interpreted as a manifestation of incoherent macroscopic quantum tunneling (MQT) of the spin.

As long as the external magnetic field $H_z$ is much smaller 
than the internal exchange interactions between the Mn ions of the 
Mn$_{12}$ cluster, the Mn$_{12}$ cluster behaves like a large single spin ${\bf S}$ of length $s=10$. 
For temperatures $T\gtrsim 1$ K its spin dynamics
can be described by a spin Hamiltonian of form $\cH=\cH_{\rm a}+\cH_{\rm Z}+\cH_{\rm sp}+\cH_{\rm T}$ including the coupling between this large spin and the phonons in the crystal
\cite{LL1Mn12,LL2Mn12,LL3Mn12,Villain,Fort,Luis,VILLAIN,GARANIN,HernandezHxTunneling}.  
In particular, 
\be 
\cH_{\rm a}=-AS_z^2-BS_z^4 
\label{H_a} 
\ee 
represents the magnetic anisotropy where $A\gg B> 0$. 
The Zeeman term through which the
external magnetic field $H_z$ couples to the spin ${\bf S}$ is given by
${\cal H}_{\rm Z}=g\mu_BH_zS_z$,
while the tunneling between  $S_z$-states is governed by
\be
{\cal H}_{\rm T}=-\frac{1}{2}B_4\left(S_+^4+S_-^4\right)+g\mu_BH_x S_x\,
,
\label{H_T}
\ee
where  $H_x=|{\bf H}|\sin\theta$ ($\ll H_z$)
is the transverse field, with $\theta$
being the misalignment angle.
The values of the anisotropy constants $A,B,B_4$ have been determined by
ESR experiments\cite{Barra,Zhong}.
Finally, the most general spin-phonon coupling reads
\bea
{\cal H}_{\rm sp}& = & g_1(\epsilon_{xx}-\epsilon_{yy})\otimes
(S_x^2-S_y^2)+
\frac{1}{2}g_2\epsilon_{xy}\otimes\{S_x,S_y\} \\
& & +\left.\frac{1}{2}g_3(\epsilon_{xz}\otimes \{S_x,S_z\}+\epsilon_{yz}
\otimes\{S_y,S_z\})+\frac{1}{2}g_4(\omega_{xz}\otimes
\{S_x,S_z\}+\omega_{yz}\otimes\{S_y,S_z\})\, , \right. \nonumber
\label{H_sp}
\eea
where $g_i$ are the spin-phonon coupling constants, and
$\epsilon_{\alpha \beta} $ ($\omega_{\alpha\beta}$) is the
(anti-)symmetric part of the strain tensor.
From the comparison between experimental data\cite{Friedman,Thomas}
and calculation it turns out that the constants $g_i\approx A$ $\forall i$
\cite{LL1Mn12,LL2Mn12,LL3Mn12}.

We denote by
$\left|m\right>$, $-s\leq m\leq s$,
the eigenstate of the unperturbed Hamiltonian
${\cal H}_{\rm a}+{\cal H}_{\rm Z}$ with eigenvalue
$\varepsilon_{m}=-Am^2-Bm^4+g\mu_BH_zm$.
If the external magnetic field $H_z$ is increased one gets doubly
degenerate spin states whenever a level $m$ coincides with a level
$m'$ on the opposite side
of the potential barrier.
The resonance condition for double degeneracy,
{\it i.e.} $\varepsilon_{m}=\varepsilon_{m'}$,  leads to the resonance field
\be
H_z^{mm'}=\frac{n}{g\mu_B}\left[A+B\left(m^2+m'^2\right)\right].
\label{condition}
\ee
As usual, we refer to $n=m+m'=$ even (odd) as even (odd) resonances.

The relaxation of the magnetization is described
in terms of a generalized master equation for the reduced density matrix $\rho(t)$
which includes off-diagonal terms due to resonances\cite{LL1Mn12,LL2Mn12}.
We use the notation $\rho_{mm'}=\left<m\left|\rho\right|m'\right>$,
$\rho_m=\left<m\left|\rho\right|m\right>$.
In the stationary limit $\dot{\rho}_{mm'}\approx 0$ a complete
master equation 
\be
\dot{\rho}_m  = -W_m\rho_m+\!\!\!\sum_{n\ne m,m'}\!\!\!
W_{mn}\rho_n\,+\Gamma_m^{m'}(\rho_{m'}-\rho_m)
\label{finalmeq}
\ee
can be derived,
where
\be
\Gamma_m^{m'}=E_{mm'}^2\frac{W_m+W_{m'}}{4\xi_{mm'}^2
+\hbar^2\left(W_m+W_{m'}\right)^2}
\label{Lorentzian}
\ee
is the incoherent tunneling rate from $m$ to $m'$
in the presence of phonon-damping\cite{LL2Mn12}. 
The spin-phonon rates $W_{m\pm 1,m}$ and $W_{m\pm 2,m}$ are evaluated
by means of Fermi's golden rule\cite{LL2Mn12}.

In Refs.~\cite{LL1Mn12,LL2Mn12} the master equation is solved
exactly to find the largest relaxation time.
The result is plotted in Fig.~\ref{overall}.
The even resonances are induced by the quartic
$B_4$-anisotropy, whereas the odd resonances are induced by
product-combinations
of $B_4S_\pm^4$- and $H_x S_x$-terms.   
In Fig.~\ref{fourpeaks} the peaks of the
resonance at $H_z=0$  (induced only by the $B_4$-term) are 
displayed
for four different temperatures. All the peaks are of single Lorentzian shape as a result of the 2-state transition
rate $\Gamma_m^{m'}$ given in (\ref{Lorentzian}), which agrees well
with the measurements\cite{Friedman}.

It is instructive to determine the dominant
transition paths via which the spin can relax.
For this an approximate analytic expression for
the relaxation time can be derived by means of conservation laws
that resemble Kirchhoff's rules for electrical circuits\cite{LL1Mn12,LL2Mn12}.\\


\hspace*{-\parindent}{\bf Incoherent Zener tunneling in Fe$_8$}\\
Besides Mn$_{12}$ there have been several experiments on the molecular magnet Fe$_8$ that
revealed macroscopic quantum tunneling of the spin\cite{Barra_Fe8,Sangregorio,Ohm,Wernsdorfer:Science,Wernsdorfer:EPL}.
In particular, recent measurements on Fe$_8$\cite{Wernsdorfer:Science,Wernsdorfer:EPL} lead to the development of the concept of the incoherent Zener tunneling\cite{LLFe8}. The resulting Zener tunneling probability $P_{\rm inc}$ exhibits Berry phase oscillations as a function of the external transverse field $H_x$.

For many physical systems the Landau-Zener model\cite{Zener} has become an important tool for studying tunneling transitions\cite{Tool1,Tool2,Tool3,Tool4}. It must be noted that all quantum systems to which the Zener model\cite{Zener} is applicable can be described by {\it pure} states and their {\it coherent} time evolution. Ref. \cite{LLFe8} generalizes the Zener theory in the sense that also the {\it incoherent} evolution of {\it mixed} states is taken into account (see also Refs. \cite{Tool1} and \cite{DZener1,DZener2,DZener3,DZener4,DZener5} for a comparison). In particular, the theory presented in Ref. \cite{LLFe8} agrees well with recent measurements of 
$P_{\rm inc}(H_x)$ for various temperatures in Fe$_8$\cite{Wernsdorfer:Science,Wernsdorfer:EPL}.

For the Zener transition usually only the asymptotic limit is of interest. Therefore it is required that the range over which $\varepsilon_{mm'}(t)=\varepsilon_m-\varepsilon_{m'}$ is swept is much larger than the tunnel splitting $E_{mm'}$ and the decoherence rate $\hbar\gamma_{mm'}$ (see below and Fig.~\ref{crossing}). In addition, the evolution of the spin system is restricted to times $t$ that are much longer than the decoherence time $\tau_{\rm d}=1/\gamma_{mm'}$. In this case, tunneling transitions between pairs of degenerate excited states are incoherent. This tunneling is only observable if the temperature $T$ is kept well below the activation energy of the potential barrier. Accordingly, one is interested only in times $t$ that are larger than the relaxation times of the excited states. Thus, the formalism presented in Refs.~\cite{LL1Mn12,LL2Mn12} can be applied.
It was shown in Ref.~\cite{LLFe8} that the Zener tunneling can be described by Eq.~(\ref{finalmeq}),
where
\be
\Gamma_m^{m'}(t)=\frac{E_{mm'}^2}{2}\frac{\gamma_{mm'}}{\varepsilon_{mm'}^2(t)
+\hbar^2\gamma_{mm'}^2}
\label{lorentzian}
\ee
is time-dependent, in contrast to Eq.~(\ref{Lorentzian}). As usual, the abbreviations $\gamma_{mm'}=(W_m+W_{m'})/2$ and $W_m=\sum_nW_{nm}$ are used, where $W_{nm}$ denotes the approximately time-independent transition rate from $\left|m\right>$ to $\left|n\right>$, which can be obtained via Fermi's golden rule\cite{LL1Mn12,LL2Mn12}. 
The tunnel splitting\cite{LL1Mn12,LL2Mn12} is given by
\be
E_{mm'}=2\left|\sum\limits_{m_1,\ldots,m_N \atop m_i\ne m,m'}
\frac{V_{m,m_1}}{\varepsilon_m-\varepsilon_{m_1}}\prod\limits_{i=1}^{N-1}
\frac{V_{m_i,m_{i+1}}}{\varepsilon_m-\varepsilon_{m_{i+1}}}V_{m_N,m'}\right|.
\label{tsplitting}
\ee
$V_{m_i,m_j}$ denote off-diagonal matrix elements of the total Hamiltonian $\cH_{\rm tot}$.  

Since all resonances $n$ lead to similar results, Eq.~(\ref{finalmeq}) is solved only in the unbiased case --- corresponding to $n=0$ (see below) --- where the ground states $\left|s\right>$, $\left|-s\right>$ and the excited states $\left|m\right>$, $\left|-m\right>$, $m\in\left[[s]-s+1,s-1\right]$ of the spin system with spin $s$ are pairwise degenerate. In addition, it is assumed that the excited states are already in their stationary state, i.e., $\dot{\rho}_m=0$ $\forall m\ne s,-s$.  
Eq.~(\ref{finalmeq}) leads then to
\be
1-P_{\rm inc}\equiv\Delta\rho(t)=\exp\left\{-\int_{t_0}^t dt'\;\Gamma_{\rm tot}(t')\right\},
\label{exact_solution}
\ee
where $\Delta\rho(t)=\rho_{s}-\rho_{-s}$, which satisfies the initial condition $\Delta\rho(t=t_0)=1$, and thus $P_{\rm inc}(t=t_0)=0$.  
The total time-dependent relaxation rate is given by
$\Gamma_{\rm tot}=2\left[\Gamma_{s}^{-s}+\Gamma_{\rm th}\right]$,
where the thermal rate $\Gamma_{\rm th}$, which determines the incoherent relaxation via the excited states, is evaluated by means of relaxation diagrams\cite{LL1Mn12,LL2Mn12}. 

Assuming linear time dependence, i.e., $\varepsilon_{mm'}(t)=\alpha_m^{m'} t$, in the transition region\cite{Zener}, and with $\left|\varepsilon_{mm'}^{<,>}\right|\gg\hbar\gamma_{mm'}$ one obtains from Eq.~(\ref{exact_solution})  
\bea
\Delta\rho
& = & \exp\left\{-\frac{2E_{s,-s}^2}{\hbar\alpha_s^{-s}}
\arctan\left(\frac{\alpha_s^{-s}}{\hbar\gamma_{s,-s}}t\right)
-\int_{-t}^t dt'\;\Gamma_{\rm th}\right\} \nn\\
& \approx & \exp\left\{-\frac{\pi E_{s,-s}^2}{\hbar\alpha_s^{-s}}
-\int_{-t}^t dt'\;\Gamma_{\rm th}\right\},
\label{rate_linear}
\eea
where $t_0=-t$.
In the low-temperature limit $T\rightarrow 0$ the excited states are not populated anymore and thus $\Gamma_{\rm th}$, which consists of intermediate rates that are weighted by Boltzmann factors $b_m$\cite{LL1Mn12,LL2Mn12}, vanishes.
Consequently, 
Eq.~(\ref{rate_linear}) simplifies to
\be
\Delta\rho=\exp\left\{-\frac{\pi E_{s,-s}^2}{\hbar\alpha_s^{-s}}\right\}=
\exp\left\{-\frac{\pi E_{s,-s}^2}{\hbar\left|\dot{\varepsilon}_{s,-s}(0)\right|}\right\}.
\label{rate_linear_limit}
\ee
The exponent in Eq.~(\ref{rate_linear_limit}) differs by a factor of 2 from the Zener exponent\cite{Zener}. This is not surprising since $\Gamma_{\rm tot}$ is the relaxation rate of $\Delta\rho$, where both $\rho_{s}$ and $\rho_{-s}$ are changed in time by the same amount, and {\it not} an escape rate like in the case of coherent Zener transition, where only the population of the inital state is changed in time.
Eq.~(\ref{rate_linear_limit}) implies $P_{\rm inc}=1$ for $\left|\dot{\varepsilon}_{s,-s}(0)\right|\rightarrow 0$ (adiabatic limit) and $P_{\rm inc}=0$ for $\left|\dot{\varepsilon}_{s,-s}(0)\right|\rightarrow\infty$ (sudden limit).

In accordance with earlier work\cite{Wernsdorfer:Science,Wernsdorfer:EPL,Barra_Fe8,Sangregorio,Ohm,Caciuffo} Ref.~\cite{LLFe8} uses a single-spin Hamiltonian $\cH=\cH_{\rm a}+\cH_{\rm T}+\cH_{\rm Z}+\cH_{\rm sp}$ that describes sufficiently well the behavior of the large spin $\bS$ with $s=10$ of a Fe$_8$ cluster.
After fitting the parameters the incoherent Zener theory is in excellent agreement with experiments\cite{Wernsdorfer:Science,Wernsdorfer:EPL} for the temperature range $0.05$ K$\le T\le 0.7$ K if the states $\left|\pm 10\right>$, $\left|\pm 9\right>$, and $\left|\pm 8\right>$ are taken into account. In particular, the path leading through $\left|\pm 8\right>$ gives a non-negligible contribution for $T\gtrsim 0.6$ K. Solving the relaxation diagram shown in Fig.~\ref{diag_Fe8} one obtains from Eq.~(\ref{rate_linear}) for Fe$_8$ in the case $n=0$
\bea
\Gamma_{\rm tot} & = & 2\left(\Gamma_{10}^{-10}+\sum\limits_{n=9}^8
\frac{b_n}{\frac{2}{W_{10,n}}+\frac{1}{\Gamma_n^{-n}}}\right), \\ 
\Delta\rho & = & \exp\left\{-\frac{\pi E_{10,-10}^2}{\hbar\alpha_{10}^{-10}}
-\sum\limits_{n=9}^8\frac{\pi E_{n,-n}^2W_{10,n}b_n}{\alpha_n^{-n}
\sqrt{E_{n,-n}^2+\hbar^2W_{10,n}^2}}\right\}, \nn
\eea
where the approximation $\gamma_{n,-n}\approx W_{10,n}$ and $\left|\varepsilon_{mm'}^{<,>}\right|\gg E_{n,-n},\gamma_{n,-n}$ is used.
$P_{\rm inc}=1-\Delta\rho$, which is plotted in Fig.~\ref{zener_Fe8}, is in good agreement with the measurements\cite{Wernsdorfer:EPL}. \\


\hspace*{-\parindent}{\bf Coherent N{\'e}el vector tunneling in 
antiferromagnetic molecular wheels}\\
Antiferromagnetic molecular clusters belong to the most promising candidates
for the observation of coherent quantum tunneling on the mesoscopic scale
currently available~\cite{chiolero:98}. Several systems in which an even 
number $N$ of
antiferromagnetically coupled ions is arranged on a ring have been
synthesized to 
date~\cite{taft:94,caneschi:96,waldmann:00,slageren:02}. 
These systems are well described by the spin Hamiltonian
\begin{equation}
\hat{H} = J \sum_{i=1}^N \hat{\bf s}_i \cdot \hat{\bf s}_{i+1} + 
g \mu_B {\bf B} \cdot  \sum_{i=1}^N \hat{\bf s}_i 
- k_z   \sum_{i=1}^N \hat{s}_{i,z}^2,
\label{eq:fewheel-h}
\end{equation}
where $\hat{\bf s}_i$ is the spin operator at site $i$ with spin quantum 
number $s$, 
$\hat{\bf s}_{N+1} \equiv \hat{\bf s}_1$, $J$ is the nearest-neighbor 
exchange, ${\bf B}$ the magnetic field, and $k_z>0$ the single-ion anisotropy
directed along the ring axis.
The parameters $J$ and $k_z$ have been well established both for 
various ferric wheels~\cite{taft:94,caneschi:96,waldmann:00,affronte:99,cornia:99,normand:00,pilawa:98,waldmann:99} with $N=6, 8, 10$, and, more 
recently, also for a Cr wheel~\cite{slageren:02}. For 
${\bf B}=0$, the classical ground-state spin configuration of the wheel shows
alternating (N{\'e}el) order with the spins pointing 
along $\pm {\bf e}_z$. The two states 
with the N{\'e}el vector ${\bf n}$ along $\pm{\bf e}_z$ [Fig.~\ref{fig:fw1}], 
labeled $|\uparrow \rangle$ and $|\downarrow \rangle$, are 
energetically degenerate and separated by an energy 
barrier of height  $N k_z s^2$. Because antiferromagnetic exchange
induces dynamics of N{\'e}el ordered spins,
the states
$|\uparrow \rangle$ and  $|\downarrow \rangle$ are not energy eigenstates. 
Rather, a molecule prepared in spin state $|\uparrow\rangle$ would tunnel
coherently between $|\uparrow \rangle$ and  $|\downarrow \rangle$ at a rate
$\Delta/h$, where $\Delta$ is the tunnel splitting~\cite{barbara:90,krive:90}. 
This tunneling of the N{\'e}el vector corresponds to a simultaneous tunneling
of all $N$ spins within the wheel through a potential barrier 
governed by the easy-axis anisotropy. Within the framework of coherent
state spin path integrals, an explicit expression for the tunnel splitting
$\Delta$ as a function of the magnetic field ${\bf B}$ has been 
derived~\cite{chiolero:98}. 
A magnetic field applied in the ring plane, $B_x$, gives rise to a Berry phase
acquired by the spins during tunneling~\cite{LLBerry,garg:93}.
The resulting interference of different tunneling paths
leads to a sinusoidal dependence of $\Delta$ on $B_x$, which allows one to
continuously tune the tunnel splitting from $0$ to a maximum value which
is of order of some Kelvin for the antiferromagnetic wheels synthesized
to date.

The tunnel splitting $\Delta$ also enters the energy spectrum of the 
antiferromagnetic wheel as level spacing between the ground and first excited 
state. Thus, $\Delta$ can be experimentally determined from
various quantities such as magnetization, static 
susceptibility, and specific heat. Even more information on the physical 
properties of antiferromagnetic wheels [Eq.~(\ref{eq:fewheel-h})] can be
obtained from a theoretical and experimental investigation of dynamical
quantities, such as the correlation functions of the total spin
$\hat{\bf S} = \sum_{i=1}^{N} \hat{\bf s}_i$ or of single spins within
the wheel~\cite{meier:01,meier:01b,honecker:01}. By 
symmetry arguments, it 
follows that the correlation function of total spin, $\langle \hat{S}_\alpha
(t) \hat{S}_\alpha (0) \rangle$, which is experimentally accessible
via measurement of the alternating current (AC) susceptibility does not contain
a component which oscillates with the tunnel frequency 
$\Delta/h$~\cite{meier:01,meier:01b,honecker:01}. Hence, neither the tunnel 
splitting nor the decoherence rate of N{\'e}el vector tunneling 
can be obtained by experimental techniques 
which couple to the {\it total} spin of the wheel. 
In contrast, the correlation function of a single spin
\begin{equation}
\langle \hat{s}_{i,z}(t) \hat{s}_{i,z} (0) \rangle \simeq s^2 \left(
\frac{e^{-\beta \Delta/2}}{2 \cosh (\beta \Delta/2)} e^{i \Delta t/\hbar}
+ \frac{e^{\beta \Delta/2}}{2 \cosh (\beta \Delta/2)} e^{-i \Delta t/\hbar}
\right)
\label{eq:neel-susc}
\end{equation}
exhibits the time dependence characteristic of coherent tunneling of the
quantity $\hat{s}_{i,z}$ with a tunneling rate $\Delta/h$~\cite{meier:01}. 
We conclude that {\it local} spin probes are required for the observation 
of the N{\'e}el vector dynamics. Nuclear spins which 
couple (predominantly) to a given single spin 
$\hat{\bf s}_i$ are ideal candidates for
such probes~\cite{meier:01} and have already been used to 
study spin cross-relaxation between electron and nuclear 
spins in ferric wheels~\cite{julien:99,cornia:00}.

For simplicity, we consider a single nuclear spin $\hat{\bf I}$, $I=1/2$,
coupled to one electron spin by a hyperfine contact
interaction $\hat{H}^\prime = A \hat{\bf s}_1 \cdot \hat{\bf I}$. According to 
Eq.~(\ref{eq:neel-susc}), the tunneling electron spin $\hat{\bf s}_1$ 
produces a rapidly oscillating hyperfine field $A s \cos (\Delta t/\hbar)$ 
at the site of the nucleus.
Signatures of the coherent electron spin tunneling can thus also be
found in the nuclear susceptibility. For 
a static magnetic field applied in the plane of the ring, $B_x$, 
it can be shown that the nuclear susceptibility
\begin{eqnarray}
&&\chi^{\prime \prime}_{I,yy}(\omega)\simeq \frac{\pi}{4} \Bigl[ \tanh \left( 
\frac{\beta \gamma_I B_x}{2}\right) \, \delta(\omega-\gamma_I B_x/\hbar) 
\label{eq:i-susc}  \\
&& + \left( \frac{A s}{\Delta} \right)^2 \! \tanh \left( 
\frac{\beta \Delta}{2} \right) \delta(\omega-\Delta/\hbar) \Bigr]
- [\omega \rightarrow - \omega] 
  \nonumber
\end{eqnarray}
exhibits a {\it satellite resonance at the tunnel splitting $\Delta$ of the 
electron spin system}~\cite{meier:01}. Here, $\gamma_I B_x$ is the Larmor 
frequency of the
nuclear spin and the first term in Eq.~(\ref{eq:i-susc}) corresponds to the 
transition between the Zeeman-split energy levels of $I$. 
Because typically $A s \simeq 1$~mK and $\Delta 
\lesssim 2$~K in Fe$_{10}$, the spectral weight of the satellite peak is 
small compared to the one of the first term in Eq.~(\ref{eq:i-susc}) unless 
the magnetic field is tuned such that  
$\Delta$ is significantly reduced compared to its maximum value. The 
observation of the satellite peak in $\chi^{\prime \prime}_{I,yy}(\omega)$
is  challenging, but possible with current experimental 
techniques~\cite{meier:01}. The experiment must be conducted with 
single crystals of an antiferromagnetic molecular wheel with sufficiently 
large anisotropy 
$k_z > 2 J/(N s)^2$ at high, tunable  fields ($10$~T) and low 
temperatures ($2$~K). 
Moreover, because the tunnel splitting 
$\Delta ({\bf B})$ depends sensitively on the relative orientation 
of ${\bf B}$ and the easy axis~\cite{cornia:99,normand:00}, careful field 
sweeps are necessary to ensure that the satellite peak in 
Eq.~(\ref{eq:i-susc}) has a large spectral weight. 

The need for local spin probes such as NMR or inelastic neutron scattering to
detect coherent N{\'e}el vector tunneling 
can be traced back to the translation symmetry of the spin
Hamiltonian $\hat{H}$~\cite{meier:01b,honecker:01}. If this symmetry is broken,
e.g. by doping of the wheel, ESR also provides an adequate technique for the
detection of coherent N{\'e}el vector tunneling. If one of the
original Fe or Cr ions of the wheel with spin $s=5/2$ or $s=3/2$, 
respectively, is replaced by an ion with different spin $s^\prime \neq s$,
this will in general also result in a different exchange constant 
$J^\prime$ and single ion anisotropy $k_z^\prime$ at the dopand site, i.e., 
\begin{equation}
 \hat{H} = J \sum_{i=2}^{N-1} \hat{{\bf s}}_i \cdot \hat{{\bf s}}_{i+1} + 
J^\prime ( \hat{{\bf s}}_1 \cdot \hat{{\bf s}}_{2} + \hat{{\bf s}}_1 
\cdot \hat{{\bf s}}_{N} )
 + g \mu_B {\bf B} \cdot  
\sum_{i=1}^N \hat{{\bf s}}_i - (k_z^\prime \hat{s}_{1,z}^2 +
k_z   \sum_{i=2}^{N} \hat{s}_{i,z}^2).
\label{eq:dwheel-h}
\end{equation}
Although thermodynamic quantities, such as magnetization, of the doped
wheel may differ significantly from the ones of the 
undoped wheel, the picture of spin tunneling in antiferromagnetic 
molecular systems~\cite{barbara:90,krive:90} remains 
valid~\cite{meier:01b}. However, due to 
unequal sublattice spins, a net total spin remains even in the 
N{\'e}el ordered state of the doped wheel [Fig.~\ref{fig:fw2}] which allows 
one to distinguish the configurations sketched in Fig.~\ref{fig:fw1} according
to their total spin. The dynamics 
of the total spin $\hat{\bf S}$ is coupled to the one of 
the N{\'e}el vector, and coherent tunneling of the 
N{\'e}el vector results in
a coherent oscillation of the total spin such that the tunneling dynamics
can also be probed by ESR. The
AC susceptibility shows a resonance peak at the tunnel splitting
$\Delta$,
\begin{equation}
\chi^{\prime \prime}_{zz} (\omega \simeq \Delta/\hbar) = \pi (g \mu_B)^2 
|\langle e|\hat{S}_z|g \rangle|^2 
\tanh \left( \frac{\beta \Delta}{2} \right) \,
\delta(\omega - \Delta/\hbar).
\label{eq:dwheel-susc}
\end{equation}
with a transition matrix element between the ground state $|g\rangle$ 
and first excited state $|e\rangle$,
\begin{equation}
|\langle e| \hat{S}_z|g \rangle| \simeq |s^\prime-s|  
\frac{8 J k_z s^2}{(g \mu_B B_x)^2}
\label{eq:dwheel-matrix}
\end{equation}
for $g \mu_B B_x \gg s \sqrt{8 J k_z}$.
The matrix element in Eq.~(\ref{eq:dwheel-matrix}) determines the spectral
weight of the absorption peak in the ESR spectrum. The analytical dependence
has been determined within a semiclassical framework and is in good 
agreement with numerical results obtained from exact diagonalization of small
systems [Fig.~\ref{fig:fw2}]. 

In conclusion, several antiferromagnetic molecular wheels
synthesized recently are promising candidates for the observation of
coherent N{\'e}el vector tunneling. Although the observation
of this phenomenon is experimentally challenging, nuclear magnetic
resonance, inelastic neutron scattering, and ESR on doped wheels are
adequate experimental techniques. The theory of coherent spin quantum tunneling
as presented above applies to zero-dimensional systems, such as 
small ferric wheels. With increasing wheel size, the possibility of 
different magnetic domains arises and new exciting quantum effects 
in the dynamics of domain walls come into play~\cite{loss:98,braun:94,braun:95,braun:96,braun:96b,braun:97,kyriakidis:98}. Molecular wheels will also allow 
one to study the transition from zero- to one-dimensional quantum behavior
with increasing system size. \\


\hspace*{-\parindent}{\bf Quantum computing with molecular magnets}\\
Shor and Grover demonstrated that a quantum computer can outperform any classical computer in factoring numbers\cite{Shor} and in searching a database\cite{Grover} by exploiting the parallelism of quantum mechanics. Recently, the latter has been successfully implemented\cite{Ahn} using Rydberg atoms. In Ref.~\cite{leuenberger:01} an implementation of Grover's algorithm was proposed that uses molecular magnets\cite{Friedman,Thomas,Wernsdorfer:EPL,Sangregorio,Thiaville}. It was shown theoretically that molecular magnets can be used to build dense and efficient memory devices based on the Grover algorithm. In particular, one single crystal can serve as a storage unit of a dynamic random access memory device. Fast electron spin resonance pulses can be used to decode and read out stored numbers of up to $10^5$, with access times as short as $10^{-10}$ seconds. This proposal should be feasible using the molecular magnets Fe$_8$ and Mn$_{12}$.

Suppose we want to find a phone number in a phone book consisting of $N=2^n$ entries. Usually it takes $N/2$ queries on average to be successful. Even if the $N$ entries were encoded binary, a classical computer would need approximately $\log_2N$ queries to find the desired phone number\cite{Grover}. But the computational parallelism provided by the superposition and interference of quantum states enables the Grover algorithm to reduce the search to one single query\cite{Grover}. This query can be implemented in terms of a unitary transformation applied to the single spin of a molecular magnet. Such molecular magnets, forming identical and largely independent units, are embedded in a single crystal so that the ensemble nature of such a crystal provides a natural amplification of the magnetic moment of a single spin. However, for the Grover algorithm to succeed, it is necessary to find ways to generate arbitrary superpositions of  spin eigenstates. For spins larger than 1/2 this turns out to be a highly non-trivial task as spin excitations induced by magnetic dipole transitions in conventional electron spin resonance (ESR) can change the magnetic quantum number $m$ by only $\pm 1$. To circumvent such physical limitations it was proposed to use multifrequency coherent magnetic radiation that allows the controlled generation of arbitrary spin superpositions. In particular, it was shown that by means of advanced ESR techniques it is possible to coherently populate and manipulate many spin states simultaneously by applying one single pulse of a magnetic a.c. field containing an appropriate number of matched frequencies. This a.c. field creates a nonlinear response of the magnet via multiphoton absorption processes involving particular sequences of $\sigma$ and $\pi$ photons which allows the encoding and, similarly, the decoding of states. Finally, the subsequent read-out of the decoded quantum state can be achieved by means of pulsed ESR techniques. These exploit the non-equidistance of energy levels which is typical of molecular magnets. 

Molecular magnets have the important advantage that they can be grown naturally as single crystals of up to 10-100 $\mu$m length containing about $10^{12}$ to $10^{15}$ (largely) independent units so that only minimal sample preparation is required. The molecular magnets are described by a single-spin Hamiltonian of the form $\cH_{\rm spin}=\cH_{\rm a}+V+\cH_{\rm sp}+ \cH_{\rm T}$\cite{LL1Mn12,LL2Mn12,LLFe8}, where $\cH_{\rm a}=-AS_z^2-BS_z^4$ represents the magnetic anisotropy ($A\gg B>0$). The Zeeman term $V=g\mu_B\bH\cdot\bS$ describes the coupling between the external magnetic field $\bH$ and the spin $\bS$ of length $s$. The calculational states are given by the $2s+1$ eigenstates   of $\cH_{\rm a}+g\mu_B H_zS_z$ with eigenenergies $\varepsilon_m=-Am^2-Bm^4+g\mu_B H_zm$, $-s\le m\le s$. The corresponding classical anisotropy potential energy $E(\theta)=-As\cos^2\theta-Bs\cos^4\theta +g\mu_BH_zs\cos\theta$, is obtained by the substitution $S_z=s\cos\theta$, where $\theta$ is the polar spherical angle. We have introduced the notation $m,m'=m-m'$. By applying a bias field $H_z$ such that $g\mu_B H_z>E_{mm'}$, tunneling can be completely suppressed and thus $\cH_{\rm T}$ can be neglected\cite{LL1Mn12,LL2Mn12,LLFe8}. For temperatures of below 1 K transitions due to spin-phonon interactions ($\cH_{\rm sp}$) can also be neglected. In this regime, the level lifetime in Fe$_8$ and Mn$_{12}$ is estimated to be about 
$\tau_{\rm d}=10^{-7}$s, limited mainly by hyperfine and/or dipolar interactions\cite{leuenberger:01}.

Since the Grover algorithm requires that all the transition probabilites are almost the same, Ref.~\cite{leuenberger:01} proposes that all the transition amplitudes between the states $\left|s\right>$ and $\left|m\right>$, $m=1,2,..., s-1$, are of the same order in perturbation $V$. This allows us to use perturbation theory. 
A different approach uses the magnetic field amplitudes to adjust the appropriate transition amplitudes\cite{LLPA}.
Both methods work only if the energy levels are not equidistant, which is typically the case in molecular magnets owing to anisotropies. In general, if we choose to work with the states $m=m_0, m_0+1,...,s-1$, where $m_0=1,2,...,s-1$, we have to go up to $n$th order in perturbation, where $n=s-m_0$ is the number of computational states used for the Grover search algorithm (see below), to obtain the first non-vanishing contribution. Fig.~\ref{spectrum_bw} shows the transitions for $s=10$ and $m_0=5$. The $n$th-order transitions correspond to the nonlinear response of the spin system to strong magnetic fields. Thus, a coherent magnetic pulse of duration $T$ is needed with a discrete frequency spectrum $\{\omega_m\}$, say, for Mn$_{12}$ between 20 and 300 GHz and a single low-frequency 0 around 100 MHz. 
The low-frequency field $\bH_z(t)=H_0(t)\cos(\omega_0t){\bf e}_z$, applied along the easy-axis, couples to the spin of the molecular magnet through the Hamiltonian
\be
V_{\rm low}=g\mu_BH_0(t)\cos(\omega_0t)S_z,
\label{pi}
\ee
where $\hbar\omega_0\ll\varepsilon_{m_0}-\varepsilon_{m_0+1}$ and ${\bf e}_z$ is the unit vector pointing along the $z$ axis. The $\pi$ photons of $V_{\rm low}$ supply the necessary energy for the resonance condition (see below). They give rise to virtual transitions with  $\Delta m=0$, that is, they do not transfer any angular momentum, see Fig.~\ref{spectrum_bw}.
The perturbation Hamiltonian for the high-frequency transitions from $\left|s\right>$ to virtual states that are just below $\left|m\right>$, $m=m_0,\ldots,s-1$, given by the transverse fields $\bH_\perp^-(t)=\sum_{m=m_0}^{s-1}H_m(t)[\cos(\omega_mt+\Phi_m){\bf e}_x-\sin(\omega_mt+\Phi_m){\bf e}_y]$, reads
\bea
V_{\rm high}(t) & = & \sum_{m=m_0}^{s-1}g\mu_BH_m(t)\left[\cos(\omega_mt+\Phi_m)S_x-\sin(\omega_mt+\Phi_m)S_y\right] \nn\\
& = & \sum_{m=m_0}^{s-1}\frac{g\mu_BH_m(t)}{2}
\left[e^{i(\omega_mt+\Phi_m)}S_++e^{-i(\omega_mt+\Phi_m)}S_-\right],
\label{sigma}
\eea
with phases $\Phi_m$ (see below), where we have introduced the unit vectors ${\bf e}_x$ and ${\bf e}_y$ pointing along the $x$ and $y$ axis, respectively. 
These transverse fields rotate clockwise and thus produce left circularly polarized $\sigma^-$ photons which induce only transitions in the left well (see 
Fig.~\ref{Figure1_L11562}). In general, absorption (emission) of $\sigma^-$ photons gives rise to $\Delta m=-1$ ($\Delta m=+1$) transitions, and vice versa in the case of  $\sigma^+$ photons. Anti-clockwise rotating magnetic fields of the form 
$\bH_\perp^+(t)=
\sum_{m=m_0}^{s-1}H_m(t)[\cos(\omega_mt+\Phi_m){\bf e}_x+\sin(\omega_mt+\Phi_m){\bf e}_y]$ can be used to induce spin transitions only in the right well (see Fig.~\ref{Figure1_L11562})). In this way, both wells can be accessed independently.

Next we calculate the quantum amplitudes for the transitions induced by the magnetic a.c. fields (see Fig.~\ref{spectrum_bw}) by evaluating the $S$-matrix perturbatively. 
The $j$th-order term of the perturbation series of the $S$-matrix in powers of the total perturbation Hamiltonian $V(t)=V_{\rm low}(t)+V_{\rm high}(t)$ is expressed by 
\bea
S_{m,s}^{(j)} & = & \left(\frac{1}{i\hbar}\right)^j\prod_{k=1}^{j-1}\int_{-\infty}^\infty
dt_k\int_{-\infty}^\infty dt_j\Theta(t_k-t_{k+1})\nn\\
& & \times U(\infty,t_1)V(t_1)U(t_1,t_2)V(t_2)\ldots V(t_j)U(t_j,-\infty),
\eea
which corresponds to the sum over all Feynman diagrams $\cF$ of order $j$, and where 
$U(t,t_0)=e^{-i(\cH_{\rm a}+g\mu_B\delta H_z)(t-t_0)/\hbar}$ is the free propagator, $\Theta(t)$ is the Heavyside function. 
The total $S$-matrix is then given by $S=\sum_{j=0}^\infty S^{(j)}$. The high-frequency virtual transition changing $m$ from $s$ to $s-1$ is induced by the frequency  $\omega_{s-1}=\omega_{s-1,s}-(n-1)\omega_0$. The other high frequencies $\omega_m$, $m=m_0,\ldots,s-2$, of the high-frequency fields $H_m$ mismatch the level separations by $\omega_0$, that is, $\hbar\omega_m=  \varepsilon_m-\varepsilon_{m+1}+\hbar\omega_0$, see Fig.~\ref{spectrum_bw}. As the levels are not equidistant, it is possible to choose the low and high frequencies in such a way that $S_{m,s}^{(j)}=0$  for $j<n$, in which case the resonance condition is not satisfied, that is, energy is not conserved. In addition, the higher-order amplitudes $|S_{m,s}^{(j)}|$ are negligible compared to $|S_{m,s}^{(j)}|$ for $j>n$. Using rectangular pulse shapes, $H_k(t)= H_k$, if $-T/2<t<T/2$, and 0 otherwise, for $k=0$ and $k\ge m_0$, one obtains ($m\ge m_0$) 
\bea
S_{m,s}^{(n)} & = & \sum_\cF\Omega_m\frac{2\pi}{i}\left(\frac{g\mu_B}{2\hbar}\right)^n
\frac{\prod_{k=m}^{s-1}H_ke^{-i\Phi_k}H_0^{m-m_0}p_{m,s}(\cF)}{(-1)^{q_\cF}q_\cF!r_s(\cF)!
\omega_0^{n-1}} \nn\\
& & \times\delta^{(T)}(\omega_{m,s}-\sum_{k=m}^{s-1}\omega_k-(m-m_0)\omega_0),
\label{n-amplitude}
\eea
where $\Omega_m=(m-m_0)!$, is the symmetry factor of the Feynman diagrams $\cF$ (see Fig.~\ref{spectrum_bw}), $q_\cF=m-m_0-r_s(\cF)$, $p_{m,s}(\cF)=\prod_{k=m}^s\left<k\left|S_z\right|k\right>^{r_k(\cF)}$, $r_k(\cF)=0,1,2,\ldots,\le m-m_0$ is the number of $\pi$ transitions directly above or below the state $\left|k\right>$, depending on the particular Feynman diagram $\cF$, and $\delta^{(T)}(\omega)=\frac{1}{2\pi}\int_{-T/2}^{+T/2}e^{i\omega t}dt
=\sin(\omega T/2)/\pi\omega$ is the delta-function of width $1/T$, ensuring overall energy conservation for $\omega T\gg 1$. The duration $T$ of the magnetic pulses must be shorter than the lifetimes $\tau_{\rm d}$ of the states $\left|m\right>$ (see Fig.~\ref{Figure1_L11562}). 

In general the magnetic field amplitudes $H_k$ must be chosen in such a way that perturbation theory is still valid and the transition probabilities are almost equal, which is required by the Grover algorithm.
In Ref.~\cite{leuenberger:01} the amplitudes $H_k$ do not differ too much from each other due to the partial destructive interference of the different transition diagrams shown in Fig.~\ref{Figure1_L11562}. 
Ref.~\cite{LLPA} shows that the transition probabilities can be increased by increasing both the magnetic field amplitudes and the detuning energies under the condition that the magnetic field amplitudes remain smaller than the detuning energies.
In this way, both high multiphoton Rabi oscillation frequencies and small quantum computation times can be attained.
This makes both methods\cite{leuenberger:01,LLPA} very robust against decoherence sources.

In order to perform the Grover algorithm, one needs the relative phases $\varphi_m$ between the transition amplitudes $S_{m,s}^{(n)}$, which is determined by $\Phi_m=\sum_{k=s-1}^{m+1}\Phi_k+\varphi_m$, where $\Phi_m$ are the relative phases between the magnetic fields $H_m(t)$.
In this way, it is possible to read-in and decode the desired phases $\Phi_m$ for each state $\left|m\right>$.
The read-out is performed by standard spectroscopy with pulsed ESR, where the circularly polarized radiation can now be incoherent because only the absorption intensity of only one pulse is needed.
Fig.~\ref{readout} shows an example where the state $\left|7\right>$ is populated within the computational basis (excluding the ground state). Then, we would observe only transitions from $\left|7\right>$ to $\left|8\right>$ at the frequency $\omega =\omega_{8,7}$ and transitions from $\left|7\right>$ to $\left|6\right>$ at $\omega=\omega_{6,7}$, which uniquely identifies the populated level, because the levels are not equidistant. 
This spectrum identifies all the populated states unambiguously. Alternatively, one could measure the magnetization of the sample with a high precision (see e.g. Ref.~\cite{Salis}). We emphasize that the entire Grover algorithm (read-in, decoding, read-out) requires three subsequent pulses each of duration $T$ with $\tau_{\rm d}>T>\omega_0^{-1}>\omega_m^{-1}>\omega_{m,m\pm 1}^{-1}$. This gives a 'clock-speed` of about 10 GHz for Mn$_{12}$, that is, the entire process of read-in, decoding, and read-out can be performed within about $10^{-10}$ s.

The proposal for implementing Grover's algorithm works not only for molecular magnets but for any electron or nuclear spin system with non-equidistant energy levels, as is shown in Ref.~\cite{LLPA} for nuclear spins in GaAs semiconductors.
Instead of storing information in the phases of the eigenstates $\left|m\right>$\cite{leuenberger:01}, in Ref.~\cite{LLPA} the eigenenergies of $\left|m\right>$ in the generalized rotating frame are used for encoding information. The decoding is performed by bringing the delocalized state $(1/\sqrt{n})\sum_m\left|m\right>$ into resonance with $\left|m\right>$ in the generalized rotating frame. Although such spin systems cannot be scaled to arbitrarily large spin $s$ --- the  larger a spin becomes, the faster it decoheres and the more classical its behavior will be --- we can use such spin systems of given $s$ to great advantage in building dense and highly efficient memory devices.

For a first test of the nonlinear response, one can irradiate the molecular magnet with an a.c. field of frequency  $\omega_{s-2,s}/2$, which gives rise to a two-photon absorption and thus to a Rabi oscillation between the states $\left|s\right>$ and $\left|s-2\right>$. For stronger magnetic fields it is in principle possible to generate superpositions of Rabi oscillations between the states $\left|s\right>$ and $\left|s-1\right>$, $\left|s\right>$ and $\left|s-2\right>$, $\left|s\right>$ and $\left|s-3\right>$, and so on (see also Ref.~\cite{LLPA}). \\
{\bf Acknowledgements}: This work has been supported by the EU
network Molnanomag, the BBW Bern, the Swiss National Science Foundation, 
DARPA, and ARO.

\begin{figure}[f]
  \begin{center}
\leavevmode
\epsfxsize=8.5cm
\epsffile{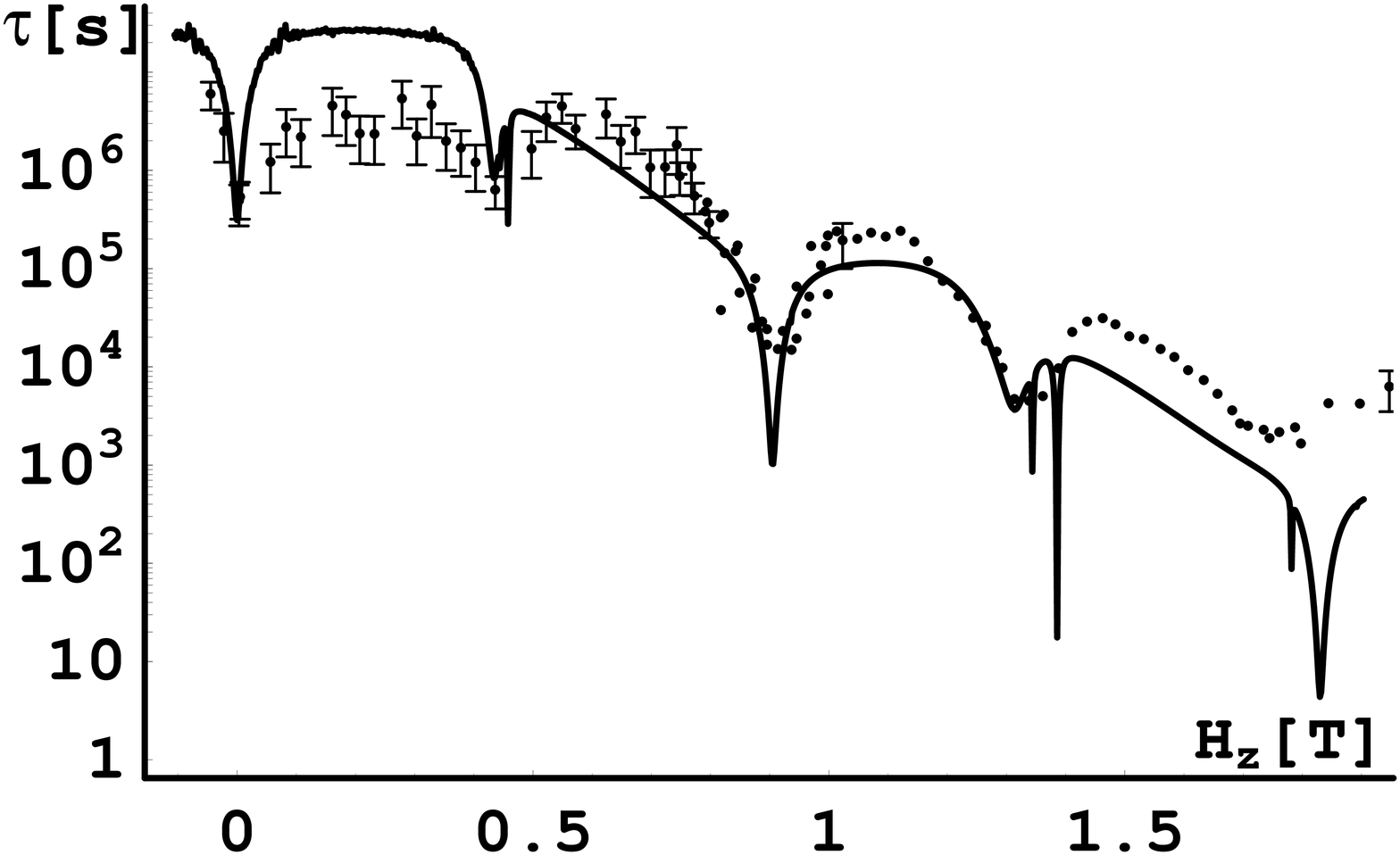}
\end{center}
\caption{Full line: semilogarithmic plot of calculated relaxation time $\tau$ as function of  magnetic field $H_z$ at $T=1.9$ K. Dots and error bars: data taken from Ref.~\protect\onlinecite{Thomas}.}
\label{overall}
\end{figure}

\begin{figure}[f]
  \begin{center}
    \leavevmode
\epsfxsize=8.5cm
\epsffile{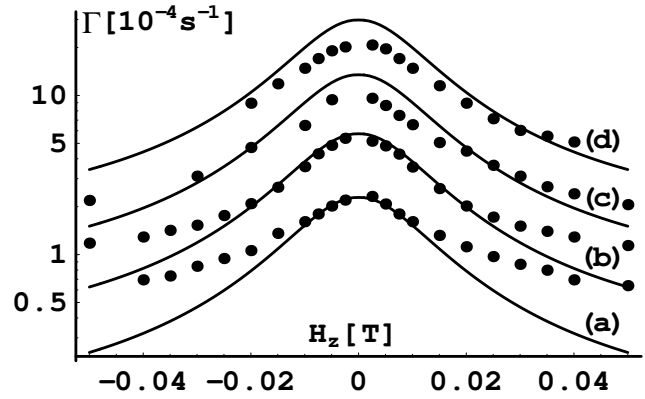}
\end{center}
\caption{Full lines: semilogarithmic plots of calculated relaxation rate $\Gamma=1/\tau$ as function of $H_z$ for the first resonance peak at (a) $T=2.5$ K, (b) $T=2.6$ K, (c) $T=2.7$ K, and (d) $T=2.8$ K. All peaks are of single Lorentzian shape. Dots: data taken from Ref.~\protect\onlinecite{Friedman}.}
\label{fourpeaks}
\end{figure}

\begin{figure}[f]
  \begin{center}
    \leavevmode
\epsfxsize=8.5cm
\epsffile{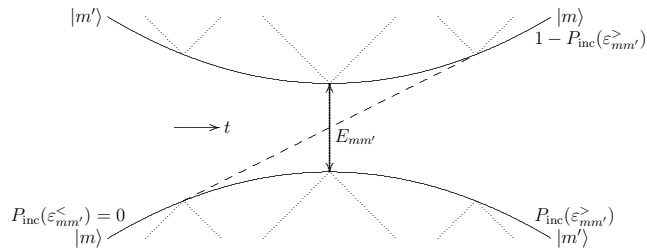}
  \end{center}
\caption{Energy level crossing diagram for incoherent Zener transitions. Dotted lines: transitions due to interaction with environment, leading to a linewidth $\gamma_{mm'}$. }
\label{crossing}
\end{figure}

\begin{figure}[f]
  \begin{center}
    \leavevmode
\epsfxsize=5cm
\epsffile{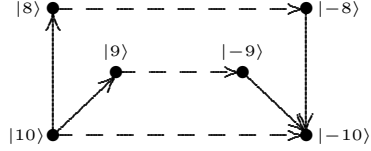}
  \end{center}
\caption{Unbiased ($n=0$) relaxation diagram for Fe$_8$. Full (dashed) lines: thermal (tunneling) transitions.}
\label{diag_Fe8}
\end{figure}

\begin{figure}[f]
  \begin{center}
    \leavevmode
\epsfxsize=8.5cm
\epsffile{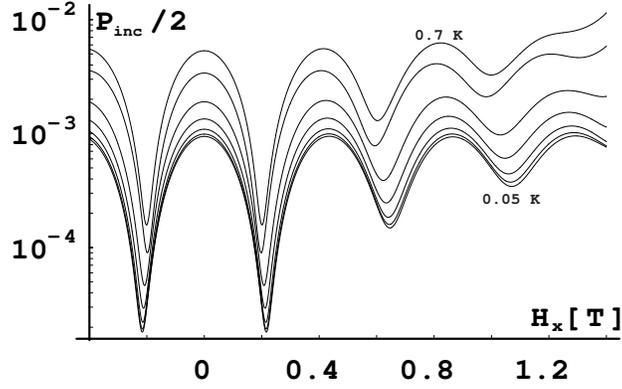}
  \end{center}
\caption{Zener transition probability $P_{\rm inc}(H_x)$ for temperatures $T=$0.7 K, 0.65 K, 0.6 K, 0.55 K, 0.5 K, 0.45 K, and 0.05 K. The fit agrees well with data (Ref.~\protect\cite{Wernsdorfer:EPL}). Note that $P_{\rm inc}$ is equal to $2P$ in Ref.~\protect\cite{Wernsdorfer:EPL}. }
\label{zener_Fe8}
\end{figure}

\begin{figure}[f]
\centerline{\psfig{file=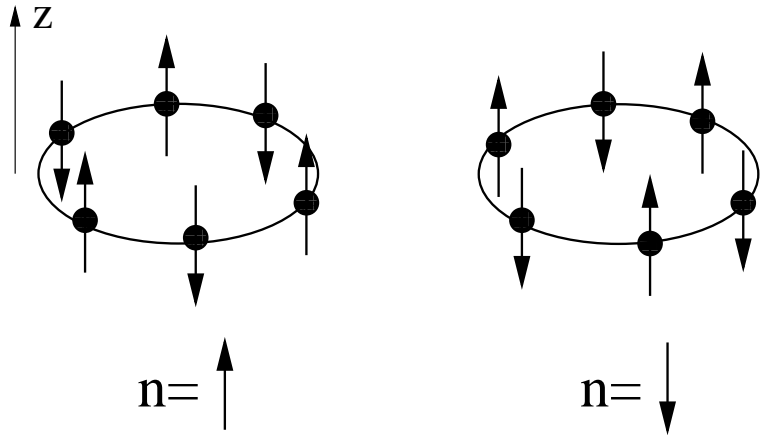,width=10cm}}
\caption{The two degenerate classical ground state spin configurations 
of an antiferromagnetic molecular wheel with easy axis anisotropy.}
\label{fig:fw1}
\end{figure} 

\begin{figure}[f]
\centerline{\psfig{file=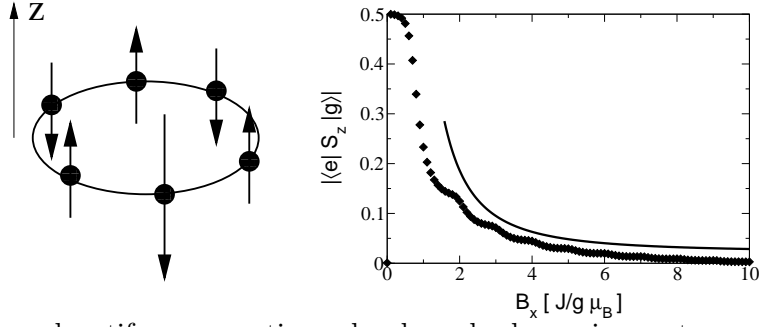,width=10cm}}
\caption{The doped antiferromagnetic molecular wheel acquires a tracer spin
which follows the N{\'e}el vector dynamics (left panel). Comparison of 
results obtained for the matrix element $|\langle e|\hat{S}_z |g\rangle|$
with a coherent state spin path integral formalism (solid line) and
by numerical exact diagonalization (symbols) for $N=4$, $s=5/2$, $s^\prime=2$,
$J^\prime=J$, $k_z=k_z^\prime=0.055$J (right panel).}
\label{fig:fw2}
\end{figure} 

\begin{figure}[f]
  \begin{center}
    \leavevmode
\epsfxsize=8.5cm
\epsffile{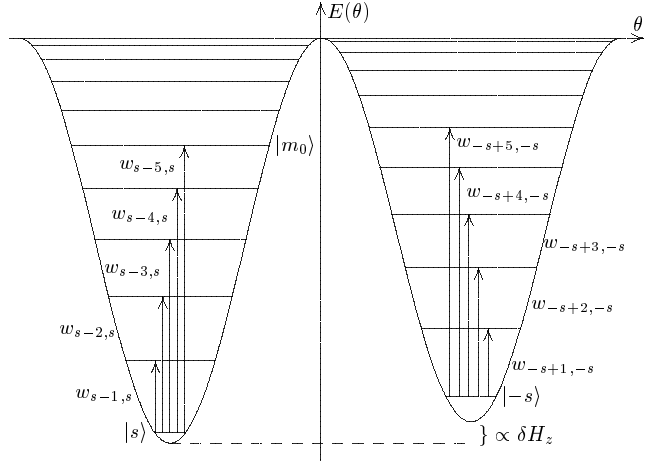}
  \end{center}
\caption{Double well potential seen by the spin due to magnetic anisotropies in Mn$_{12}$. Arrows depict transitions between spin eigenstates driven by the external magnetic field $\bH$.}
\label{Figure1_L11562}
\end{figure}

\begin{figure}[htb]
  \begin{center}
    \leavevmode
\epsfxsize=6cm
\epsffile{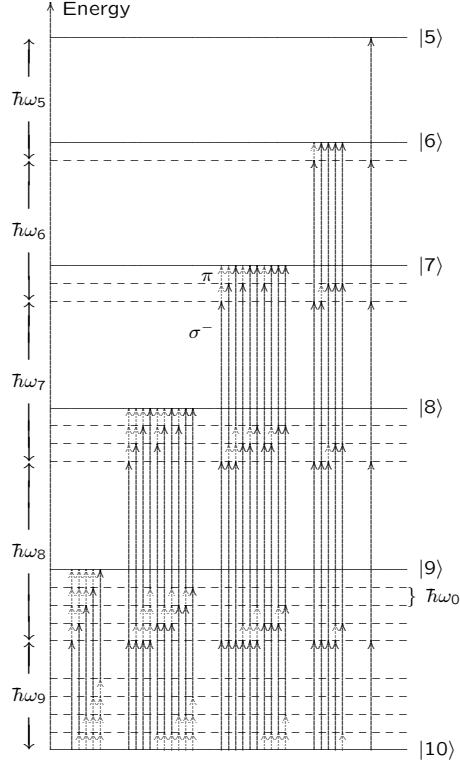}
  \end{center}
\caption{Feynman diagrams $\cF$ that contribute to $S_{m,s}^{(5)}$ for $s=10$ and $m_0=5$ describing transitions (of 5th order in $V$) in the left well of the spin system (see Fig.~\ref{Figure1_L11562}). The solid and dotted arrows indicate 
$\sigma^-$ and $\pi$ transitions governed by Eq.~(\ref{sigma}) and Eq.~(\ref{pi}), respectively.  We note that $S_{m,s}^{(j)}=0$ for $j<n$, and $S_{m,s}^{(j)}\ll S_{m,s}^{(n)}$ for $j>n$.}
\label{spectrum_bw}
\end{figure}

\begin{figure}[f]
  \begin{center}
    \leavevmode
\epsfxsize=6cm
\epsffile{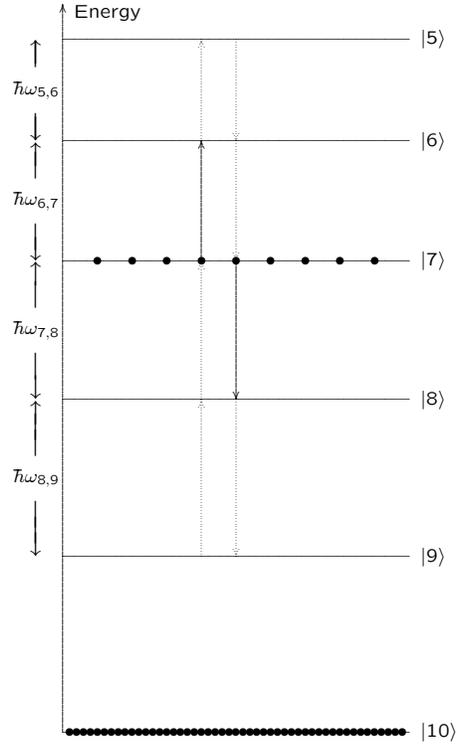}
  \end{center}
\caption{Read-out of the information by ESR in the linear regime. In this example only the state $\left|7\right>$ of the computational basis is populated. Thus only the transitions marked by solid arrows induce a response that can be observed in a ESR spectrum.}
\label{readout}
\end{figure}

\end{document}